\documentclass[aps,prd,email,preprint,showpacs,showkeys,preprintnumbers,amsmath,amssymb,nofootinbib]{revtex4}

\usepackage{color}
\usepackage{latexsym}
\usepackage{amsmath}
\usepackage{amssymb}
\usepackage{eufrak}
\usepackage{euscript}
\usepackage{pstricks}
\usepackage{graphics}
\usepackage{graphicx}
\usepackage{picture}
\def\[{\left\lbrack}
\def\]{\right\rbrack}

\def\({\left(}
\def\){\right)}

\newcommand{\bbe}{\begin{equation}}
\newcommand{\eee}{\end{equation}}
\newcommand{\eea}{\end{eqnarray}}
\newcommand{\bba}{\begin{eqnarray}}

\def\ni{\noindent}

\begin{document}

\pagestyle{myheadings}
\markright{Duality and gauge invariance of noncommutative spacetimes Podolsky....}

\title{\Large{Duality and gauge invariance of noncommutative spacetime Podolsky electromagnetic theory}}

\author{Everton M. C. Abreu$^{a,b}$}
\email{evertonabreu@ufrrj.br}
\author{Rafael L. Fernandes$^{b}$}
\email{rlfernandes@fisica.ufjf.br}
\author{Jorge Ananias Neto$^{b}$}
\email{jorge@fisica.ufjf.br}

\affiliation{$^a$Grupo de F\' isica Te\'orica e Matem\'atica F\' isica, Departamento de F\'{\i}sica, \\
Universidade Federal Rural do Rio de Janeiro\\
BR 465-07, 23890-971, Serop\'edica, RJ, Brazil \\\\ 
$^b$Departamento de F\'{\i}sica, ICE, Universidade Federal de Juiz de Fora,\\
36036-330, Juiz de Fora, MG, Brazil\\\\
\today\\}

\begin{abstract}
\noindent The interest in higher derivatives field theories has its origin mainly in their influence concerning the renormalization properties of physical models and to remove ultraviolet divergences.
In this letter we have introduced the noncommutative (NC) version of the Podolsky theory and we investigated the effect of the noncommutativity over its original gauge invariance property.  We have demonstrated precisely that the noncommutativity spoiled the gauge invariance of the original action.  After that we have used the Noether dualization technique to obtain a dual and gauge invariant action.  More than to obtain the NC Podolsky theory, we have another motivation in this work, 
which is to show that, although the introduction of noncommutativity spoils the gauge invariance, it is possible to recover this property using a standard dualization method 
which did not need any modification due to any NC effect in the original theory, by the way

\end{abstract}


\maketitle

\pagestyle{myheadings}
\markright{Duality and gauge invariance of noncommutative spacetimes Podolsky....}


It is an ordinary fact that we have an infinite electromagnetic mass in classical electrodynamics and the consequence is that it causes infinities which are associated with a point particle \cite{kruglov}.  However, this point particle limit can be achieved with a proper cutoff.  

We know that Podolsky's electrodynamics is a finite theory, namely, the electromagnetic energy associated to a point particle is finite.  When the considered distances are much greater than a cutoff, the Maxwell electrodynamics is recovered.  Besides, one can connect the Podolsky theory with the Abraham-Lorentz equation \cite{kruglov}.  Another motivation to study higher order Lagrangians is that they can be considered as effective theories on the QCD infrared sector \cite{aab} where it enforces a very nice asymptotic behavior of the gluon propagator.  Higher derivatives terms in field theory of supersymmetric fields act as a regularization mechanism \cite{bpz}.

On the other hand, one of the great challenges in theoretical physics is to unify in a single and consistent framework both theories of quantum mechanics and general relativity.  The combination of special relativity and quantum field theory was already accomplished through the Klein-Gordon and Dirac approaches.  However, the path to formulate a general relativity with the quantum mechanical approach is still a mystery so far.  

This so-called quantization procedure of general relativity had stumbled in another theoretical physics challenge, i.e., the infinities (divergences) that appear in specific calculations during the quantization process.  This question is directly connected to the understanding of the quantum fields behavior at high energy scale which is also connected to the structure of spacetime at (or near) the Planck scale.   This structure definition is necessary in order to construct the Hilbert space inner product essential to the definition of the particle concept.  There are several formalisms that deal with these questions and one of those is noncommutative (NC) geometry. 

For example, one attempt to free us from these infinities was made by Snyder \cite{snyder} when he had constructed a five dimensional NC algebra.  Unfortunately, little time after Snyder's effort, Yang \cite{yang} demonstrated that even in this NC algebra, the divergences persisted.

This result condemned Snyder noncommutativity to be outcast for more than fifty years until Seiberg and Witten \cite{sw} demonstrated that the algebra resulting from string theory embedded in a magnetic field showed itself to have a NC algebra.  Since then we have seen an industrial research production concerning several NC formulations that deserves our attention and investigation.

In a suitable basis, the algebra underlying Snyder's spacetime may be presented as
a modification of the phase space canonical commutation relations given by
\bba
\label{azero}
\[x^{\mu}, x^{\nu} \] &=& i\,l^2\,\hbar^{-1}\,\(x^{\mu}\,p^{\nu}\,-\,x^{\nu}\,p^{\mu} \) \nonumber \\
\[x^{\mu}, p_{\nu} \] &=& i\,\hbar\,\delta^{\mu}_{\,\,\,\nu}\,+\,i\,l^2\,\hbar^{-1}\,p^{\mu}\,p_{\nu} \\
\[p_{\mu}, p_{\nu} \]&=&0 \,\,.\nonumber
\eea
This algebra involves a fundamental minimal length $l$, the scale of noncommutativity, 
such that the ``classical" phase space of quantum mechanics is recovered
at $l = 0$. The original motivation behind these relations was that the introduction of the
length scale $l$ is tantamount to regard hadrons in quantum field theory as extended
objects, because at the time, renormalization theory was regarded as a distasteful procedure \cite{snyder}. 
The commutation relations (\ref{azero}) describe a discrete spacetime which is compatible with Lorentz invariance.

This theoretical framework is called NC field theory and it may be a relevant physical model at scales in between $l_P \;(\simeq 1.6 \times 10^{-33} cm)$ and $l_{LHC}\; (\simeq 2 \times 10^{-18} cm)$. In
fact, one of the main threads of research in this field has been related to studies of
energetic cosmic rays, for instance. These field theories provide fruitful avenues of
exploration for several reasons.  Firstly,
some quantum field theories are better behaved on NC spacetimes than
on ordinary spacetimes. In fact, some are completely finite, even non-perturbatively. In
this manner spacetime noncommutativity presents an alternative to supersymmetry or
string theory. Secondly, it is a useful arena for studying physics beyond the standard
model, and also for standard physics in strong external fields. Thirdly, it sheds light on
alternative lines of attack to addressing various fundamental issues in quantum field
theory, for instance the renormalization and axiomatic programs. Finally, it naturally relates field theory to gravity. Since the field theory may be easier to quantize,
this may provide significant insights into the problem of quantizing gravity.

However, in his approach, Snyder postulated an identity between coordinates and generators of the $SO(4,1)$ algebra.   Hence, he promoted the spacetime coordinates to Hermitian operators.

One of the most popular approaches of NC theories \cite{reviews} is the one governed by the well known Moyal-Weyl product.  In this approach, the standard product of two NC objects is substituted by the so-called star-product given by

\bbe
\label{A}
\hat{f}(x) * \hat{g}(x)\,=\, exp\Big(\frac i2 \theta^{\mu\nu} \partial_{\mu}^x \partial_{\nu}^y \Big)\,\hat{f}(x) \hat{g}(y) \Big|_{x=y}
\eee

\noindent where  $\theta^{\mu\nu}$ is the well known NC parameter that is present in the beginning point of the NC definition, namely,

\bbe
\label{B}
\[ \hat{x}^{\mu}, \hat{x}^{\nu} \]\,=\,i\,\theta^{\mu\nu}\,\,,
\eee

\noindent where the commutator indicates operators in Snyder's formalism.  The constant feature of $\theta^{\mu\nu}$ brings another problem, since we have a fixed direction defined in the r. h. s. of Eq. (\ref{B}).  This fact breaks the Lorentz invariance of the theory.
The solution of this problem lead us to another formulation of a NC theory formulated by Doplicher, Fredenhagen and Roberts (DFR) \cite{dfr}, which is based in general relativity and quantum mechanics arguments.

In this work we have investigated two aspects of the gauge invariance of NC theories.  The first one is if the NC extra terms spoils the gauge invariance of the original commutative theory at an ``unrecovered" level and we can also ask if we have to modify the dualization formalism due to any NC effect.   Firstly we have obtained a NC version of a high derivative model, i.e., the Podolsky electromagnetic theory, which is a new result.  The dualization method used here is the Noether formalism \cite{iw,iw2,ainrw,binrw} which was used also in the soldering formalism analysis \cite{abw,abw2,bw}.  The physical and Hamiltonian analysis of this NC Podolsky model is out of the scope of this work but it is the object of an ongoing research.


\bigskip

In a seminal paper \cite{dj}, Deser and Jackiw used the master action concept to show the dynamical equivalence between the self-dual (SD) and Maxwell-Chern-Simmons (MCS) theories.  The authors demonstrated precisely the existence of a hidden symmetry in the self-dual model.  After this result, the master action procedure was used to disclose the physics behind planar physics phenomena and bosonization, only to mention a few.  Concerning the last one, it is worth to explain that it is a procedure that expresses a theory of interacting fermions in terms of free bosons.  In $D=2$ this approach reveals the dual equivalence feature of these representations \cite{dns,dns2,dns3,dns4,dns5} being extended to higher dimensions \cite{luscher,marino}.

Motivated by this equivalence between SD and MCS models it was natural to ask if there is another way to obtain analogous and new results.  In papers analyzing the existence of hidden gauge symmetries inside second-class systems, it was proved that non-invariant models (second-class) are in fact equivalent to gauge invariant systems (first-class).  This result happens under certain gauge fixing conditions.  The main advantage of having a gauge theory is the fact that we can establish chains of equivalence between different models through different gauge fixing conditions.

The Noether embedding technique \cite{iw} is based on the traditional idea of a local lifting of a global symmetry and may be realized by an iterative embedding of Noether counter terms.  This technique was originally explored in the soldering formalism context \cite{abw,abw2,bw} and was explored in \cite{iw2,ainrw,binrw} since it seems to be the most appropriate technique for non-Abelian generalization of the dual mapping concept.

The method is an iterative procedure and the first step entails the imposition of a trivial local gauge transformation concerning the zeroth-Lagrangian, which is how we will call the original and so far untouched Lagrangian.  Of course this zeroth-Lagrangian is not gauge invariant under this imposed gauge transformation.  The computation of its variation permits us to construct the Noether currents.  So, the variation of the zeroth-Lagrangian can be written as a sum of the Noether currents coupled with the local gauge parameter.  This basically closes the first iteration. 

The second one begins with the introduction of auxiliary fields interacting with the Noether currents computed during the first iteration.  Hence, the now first-Lagrangian can be written as a sum of the zeroth-Lagrangian plus the terms with the coupling of Noether currents with the auxiliary fields.  We perform the variation of this first-Lagrangian to verify if it is gauge invariant.  If it is not gauge invariant yet, we have to introduce auxiliary fields (not necessarily new ones) again in order to try to obtain gauge invariance.   If we did not obtain that, a new Lagrangian has to be constructed and the process continues until we have gauge invariance.  The interested reader that is not pleased with this brief explanation can find more details and applications in \cite{iw,iw2,ainrw,binrw}.

We have to add an important comment here.  The Noether embedding algorithm can be applied to Abelian and non-Abelian theories independent of their dimensions.  Hence, it is a perfect procedure to work with field theory with extra dimensions, which is our case here.   We will demonstrate here that it can be applied in a more generalized way, namely, that it can be applied to commutative and NC theories.

\bigskip


Hence, let us begin with the action of NC Podolsky field which is given by

\bbe
\label{1}
S\,=\,\int \Big(-\,\frac 14 \hat{F}_{\mu\nu} * \hat{F}^{\mu\nu}\,-\,a^2\,\partial_{\lambda}\hat{F}^{\lambda\nu} * \partial^{\mu}\hat{F}_{\mu\nu} \Big)\,d^4 x
\eee

\ni where, as explained above, $*$ means the Moyal-Weyl product.  The hat notation indicates NC objects and $\hat{A}_{\mu}$ and $\hat{F}_{\mu\nu}$ are the NC vector potential and field strength tensor respectively.  The term $a$ represents $1/m$ which is the inverse mass of the $\hat{A}_{\mu}$ field.

Since we know that for two objects within an integral \cite{rsh} the Moyal-Weyl product can be substituted by ordinary product, we can write Eq. (\ref{1}) as

\bbe
\label{2}
S\,=\,\int \Big(-\,\frac 14 \hat{F}_{\mu\nu}  \hat{F}^{\mu\nu}\,-\,{a^2}\,\partial_{\lambda}\hat{F}^{\lambda\nu}  \partial^{\mu}\hat{F}_{\mu\nu} \Big)\,d^4 x
\eee

\ni and the action above describes the Podolsky theory in a NC spacetime.  In order to show the NC contribution to the respective commutative theory we have to use the well known Seiberg-Witten map \cite{sw} which tells that

\bbe
\label{3}
\hat{A}_{\mu}\,=\,A_{\mu}\,-\,\frac 12 \theta^{\alpha\beta} A_{\alpha}\,(\partial_{\beta} A_{\mu}\,+\,F_{\beta\mu})
\eee

\ni and consequently

\bbe
\label{4}
\hat{F}_{\mu\nu}\,=\,F_{\mu\nu}\,+\,\theta^{\rho\sigma}\,(F_{\mu\rho} F_{\nu\sigma}\,-\,A_{\rho} \partial_{\sigma} F_{\mu\nu} )\,\,,
\eee

\ni where $\theta^{\mu\nu}$ is the NC parameter explained above.  Using Eqs. (\ref{3}) and (\ref{4}) into Eq. (\ref{2}) we have that the NC Podolsky Lagrangian density with the same degrees of freedom and with additional terms due to the NC parameter first-order is

\bba
\label{5}
\hat{\cal L}\,&=&\, -\,\frac 14\,F^2\,-\,\theta^{\rho\sigma}\,\Big(2 F^{\mu}_{\:\:\:\rho} F^{\nu}_{\:\:\:\sigma} F_{\mu\nu}\,-\,\frac 12 F^2 F_{\rho\sigma} \Big) \nonumber \\
\,&-&\,{a^2}\,\partial_{\lambda}\,F^{\lambda\nu} \Big[\partial^{\mu} F_{\mu\nu}\,+\,2 \theta^{\rho\sigma} \partial^{\mu} \Big(F_{\mu\rho} F_{\nu\sigma} \,-\, A_{\rho} \partial_{\sigma} F_{\mu\nu} \Big) \Big] \,\,,
\eea

\ni where we can see quickly that when $\theta^{\mu\nu}=0$ we can recover the commutative theory.  The construction of the corresponding Hamiltonian analysis of Eq. (\ref{5}) is object of an ongoing research since here we are dealing with gauge invariance of (\ref{5}), which will be described just below.

\bigskip

Hence, let us use this formalism to obtain a dual action to the one in (\ref{5}) which is gauge invariant since we will show that the NC terms spoils the Podolsky gauge invariance.

Let us begin, following the dualization technique, by fixing a local gauge symmetry for the vector potential such as 

\bbe
\label{6}
\delta\,A_{\mu}\,=\,\partial_{\mu}\,\varepsilon
\eee

\ni where $\varepsilon$ is the gauge parameter.  It is trivial to see that the Maxwell field strength tensor is invariant under (\ref{6}) such as $\delta F_{\mu\nu}=\partial_{\mu} \delta A_{\nu}\,-\,\partial_{\nu} \delta A_{\mu}\,=\,\partial_{\mu} \partial_{\nu} \varepsilon\,-\,\partial_{\nu} \partial_{\mu} \varepsilon=0$.   So, in (\ref{5}), the gauge invariance is destroyed thanks to the last term.  Hence, let us focus on it

\bba
\label{7}
\hat{\cal L}_0&=&-\,{a^2}\,\partial_{\lambda} F^{\lambda\nu} \Big[ -\,2 \theta^{\rho\sigma} \partial^{\mu} \Big( A_{\rho}\partial_{\sigma} F_{\mu\nu} \Big) \Big] \nonumber \\
&=&{a^2} \theta^{\rho\sigma} \partial_{\lambda} F^{\lambda\nu} \Big[ \partial^{\mu}A_{\rho} \partial_{\sigma} F_{\mu\nu}\,+\,A_{\rho} \partial^{\mu}\partial_{\sigma} F_{\mu\nu} \Big]
\eea

\ni and the gauge variation is, after an integration by parts,

\bbe
\label{8}
\delta \hat{\cal L}_0\,=\,-\,{a^2} \theta^{\rho\sigma} \partial^{\mu}\partial_{\lambda} F^{\lambda\beta} \partial_{\sigma} F_{\mu\beta} \partial_{\rho} \varepsilon
\eee

\ni which can be written as

\bbe
\label{9}
\delta \hat{\cal L}_0\,=\,\hat{J}^{\rho} \partial_{\rho} \varepsilon\,\,.
\eee

\ni This result shows, as we claimed before that the NC Podolsky theory is not gauge invariant.  The Noether current $J^{\mu}$ is given by

\bbe
\label{10}
\hat{J}^{\rho}\,=\,-\,{a^2} \theta^{\rho\sigma} \partial^{\mu} \partial_{\lambda} F^{\lambda\nu} \partial_{\sigma} F_{\mu\nu}
\eee

\ni The next step is to construct $\hat{\cal L}_{1}$ as

\bbe
\label{11}
\hat{\cal L}_1\,=\,\hat{\cal L}_0\,-\,\hat{J}^{\mu} B_{\mu}
\eee

\ni where $B_{\mu}$ is an auxiliary field, which can be NC or not.  The gauge variation of $\hat{\cal L}_{1}$ is

\bbe
\label{12}
\delta \hat{\cal L}_1\,=\,\delta \hat{\cal L}_0\,-\,(\delta \hat{J}^{\mu}) B_{\mu}\,-\,\hat{J}^{\mu} (\delta B_{\mu})
\eee

\ni and, following the dualization process, we can choose conveniently that

\bbe
\label{13}
\delta B_{\mu}\,=\,\partial_{\mu} \varepsilon
\eee

\ni and substituting this in Eq. (\ref{12}) we have that 

\bbe
\label{14}
\delta \hat{\cal L}_1\,=\,\delta \hat{\cal L}_0\,-\,(\delta \hat{J}^{\mu}) B_{\mu}\,-\,\hat{J}^{\mu} (\partial_{\mu} \varepsilon)\,\,,
\eee

\ni where we can see that the first and the last terms in the r.h.s. of (\ref{14}) cancel out.  So,

\bbe
\label{15}
\delta \hat{\cal L}_1\,=\,-\,(\delta \hat{J}^{\mu}) B_{\mu}\,\,.
\eee

\ni However, from (\ref{10}) we can easily obtain that $\delta \hat{J}^{\mu}=0$ which makes $\delta \hat{\cal L}_1 =0$ and we have that $\hat{\cal L}_1$ is gauge invariant and it can be written as,

\bba
\label{16}
& &\hat{\cal L}_1\,=\,\hat{\cal L}_0\,-\,\hat{J}^{\mu} B_{\mu} \nonumber \\
&\Longrightarrow&\quad \hat{\cal L}_1\,=\, -\,\frac 14 F^2\,-\,\theta^{\rho\sigma} \Big(2 F^{\mu}_{\:\:\:\rho} F^{\nu}_{\:\:\:\sigma} F_{\mu\nu}\,-\,\frac 12 F^2 F_{\rho\sigma} \Big)  \\
&-& {a^2}\,\Big\{ \partial_{\lambda} F^{\lambda\nu} \Big[ \partial^{\mu} F_{\mu\nu}\,+\,2 \theta^{\rho\sigma} \partial^{\mu} 
\Big(F_{\mu\rho} F_{\nu\sigma}\,-\,A_{\rho} \partial_{\sigma} F_{\mu\nu} \Big) \,-\,\theta^{\rho\sigma} \partial^{\mu} \Big(\partial_{\sigma} F_{\mu\nu} B_{\rho} \Big) \Big] \Big\} \nonumber
\eea

\ni which is the dual action of the action in (\ref{5}) and it has the same physical properties as the original action $\hat{\cal L}_0$.  The next step would be to eliminate $B_{\mu}$ through the equations of motion for $F_{\mu\nu}$.  However, it is easy to realize that, looking at $\hat{\cal L}_1$ in (\ref{16}), it will turn $\hat{\cal L}_1$ into a huge Lagrangian with too many terms.  Hence, to simplify let us keep the form with the auxiliary field.  Since, our objective here is obtain, besides a NC version of Podolsky theory, a dual gauge invariant NC expression for the NC original action.  The auxiliary field $B_{\mu}$, although it recovers the gauge invariance of $\hat{\cal L}_0$ it cannot be considered a type of Stueckelberg field since it did not introduce a mass term.  Although in $\hat{\cal L}_1$ we have one more degree of freedom than $\hat{\cal L}_0$, as we have explained, $B_{\mu}$ can be eliminated through $F_{\mu\nu}$ equations of motion.  So, the final Lagrangian would have the same number of degrees of freedom as the original NC one.

Since $\hat{\cal L}_1$ is gauge invariant, in Dirac's language, it is a first-class system, which depicts in a general manner the physical features of constrained dynamical systems.  Since the original action $\hat{\cal L}_0$ is not gauge invariant, we can say that it is a second-class system and the conversion into a first-class results in a gauge invariant action which also has the same physical properties as the original one.  Having said all that, we can ask if there is a connection between $\hat{\cal L}_{0}$ (second-class) and $\hat{\cal L}_1$.

\bigskip


To conclude, in this work we have introduced two new results, the first one is the NC version of the Podolsky electromagnetic theory through Moyal-Weyl product and Seiberg-Witten map and, after the demonstration that the NC terms spoiled the gauge invariance, we have obtained a NC gauge invariant Podolsky theory.   Although the final action has one more degree of freedom, an auxiliary field $B_{\mu}$, it can be eliminated through $F_{\mu\nu}$ equations of motion.  Since we have not a new mass term, this auxiliary field is not a Stueckelberg-type one.

Since we have obtained a gauge invariance theory, it can be considered a first-class system following Dirac's formalism and the quantization would be through Poisson brackets.

As a perspective we can analyze the constrained system of the original NC theory since it is a second-class and convert this second-class NC system  into first-class one and verify if there is some connection between both actions the one obtained here and the converted one.   This is an ongoing research.

As another perspective we can investigate the physical (electromagnetic) properties, described at the beginning of this letter, of the NC Podolsky model and its dual in order to obtain their connections between each other and with the commutative Podolsky theory.

\section{Acknowledgments}

\ni EMCA would like to thank CNPq (Conselho Nacional de Desenvolvimento Cient\' ifico e Tecnol\'ogico), Brazilian scientific support agency, for partial financial support.

\bigskip
\bigskip


\end{document}